\documentclass[a4paper,fleqn]{cas-sc}
\usepackage[authoryear,longnamesfirst]{natbib}
\usepackage{apalike}
\usepackage[ruled,linesnumbered]{algorithm2e} 
\usepackage{subcaption}  
\usepackage{epstopdf}    
\usepackage{rotating}
\usepackage{tablefootnote}
\usepackage{multirow}
\usepackage{booktabs}
\usepackage{siunitx}
\usepackage{caption}
\usepackage{hyperref}
\usepackage{booktabs} 
\usepackage{longtable}
\usepackage{bm}
\usepackage{times} 
\usepackage{caption}
\usepackage{hyperref}
\hypersetup{
    colorlinks=true,
    citecolor=blue,
    linkcolor=blue,
    urlcolor=blue
}
\def\tsc#1{\csdef{#1}{\textsc{\lowercase{#1}}\xspace}}
\tsc{WGM}
\tsc{QE}

\begin{document}
\let\WriteBookmarks\relax
\def\floatpagepagefraction{1}
\def\textpagefraction{.001}
\let\printorcid\relax    

\shorttitle{Navigating the Congestion Maze: Geospatial Analysis and Travel Behavior Insights for Dockless Bike-Sharing Systems in Xiamen}    
\shortauthors{Xuxilu, Zhang}  

\title [mode = title]{Navigating the Congestion Maze: Geospatial Analysis and Travel Behavior Insights for Dockless Bike-Sharing Systems in Xiamen}  

\author[1]{Xuxilu, Zhang}

\fnmark[1]

\ead{zhangxuxilu814@gmail.com}


\credit{Writing – review \& editing, Writing – original
draft, Conceptualization, Methodology, Formal analysis, Data curation, Software}

\affiliation[1]{organization={School of Statistics and Data Science, Jiangxi University of Finance and Economics},
                city={Nanchang},
                postcode={330013}, 
                country={China}}


\cormark[1]

\cortext[1]{Xuxilu Zhang}

\author[2]{Lingqi Gu}
\fnmark[2]
\ead{lingqi.gu@fjnu.edu.cn}
\affiliation[2]{organization={School of Mathematics and Statistics, Fujian Normal University},
                city={Fuzhou},
                postcode={350000}, 
                country={China}}
\credit{Writing – review \& editing, Writing – original
draft, Methodology, Supervision, Funding acquisition}

\author[3]{Nan Zhao}
\fnmark[3]
\ead{jomerkar@163.com}
\affiliation[3]{organization={Minjiang University},
                city={Fuzhou},
                postcode={350000}, 
                country={China}}
\credit{Writing – review \& editing, Writing – original
draft, Data curation}

\begin{abstract}
    Shared bicycles have emerged as a transformative force in urban transportation, effectively addressing the perennial 'last mile' challenge faced by commuters. The limitations of station-based bike-sharing systems, constrained by point-to-point travel, have spurred the popularity of the dockless model, offering flexible rentals and eliminating docking infrastructure constraints. However, the rapid growth of the sharing economy has introduced new challenges, notably an imbalance between supply and demand, leading to issues like the unavailability of bicycles and insufficient parking spaces during peak hours. To address these challenges, this study introduces a novel variable, Congestion Density ($C$), to quantitatively measure dynamic congestion levels in dockless bicycle-sharing systems. Leveraging real-time shared bike information from Xiamen, China, we present a sophisticated clustering framework for congested spots, identifying 563 congested spots categorized into Over-crowded, Semi-crowded, and Light-crowded clusters. Strikingly, these clusters align with established subway lines and bus stops, revealing a prevalent trend of integration between subway/bus services and bike-sharing. Overall, this study proposes parking lot management plans and policy recommendations based on the dynamics of crowded parking spaces, geographical characteristics, and land functional attributes. Our findings provide crucial insights for implementing bike-sharing electric fences and understanding urban mobility patterns, contributing to sustainable urban transportation. 
\end{abstract}


\begin{keywords}
 \sep Dockless bike-sharing \sep Travel Behavior \sep BS congested spots \sep Supply-demand dynamics \sep K-means algorithm
\end{keywords}

\label{sec:title}
\maketitle

\section{Introduction}\label{sec:intro}

The process of urbanization has resulted in numerous challenges, with traffic congestion and pollution being pervasive issues in contemporary urban environments \citep{pucher2005urban}. As traditional transportation systems struggle to meet the escalating demand for urban mobility, innovative solutions are required \citep{laporte2018shared}. Shared bicycles have emerged as a promising remedy, offering users the flexibility for short urban journeys while potentially alleviating congestion and reducing emissions. Currently, bike-sharing systems (BSS) predominantly operate under two models: station-based bike-sharing (docked) and free-floating bike-sharing (dockless) \citep{liu2018static}.

In the station-based model, users rent bicycles from designated stations and return them to predefined slots within these stations upon completing their rides. However, this structure inherently limits point-to-point travel options, leading to operational challenges such as mismatches between ride demand and bicycle/parking availability at stations \citep{vogel2011understanding}. Conversely, the dockless model utilizes internet connectivity and mobile apps for seamless bicycle rentals, allowing users to park bikes within specified areas without docking infrastructure constraints \citep{chen2020dockless}. This enhances last-mile connectivity and door-to-door convenience, fostering integration with public transportation networks \citep{oeschger2020micromobility}.

However, the integration of free-floating bike-sharing systems (FFBSS) in the evolving sharing economy introduces managerial complexities, including user parking behaviors that lead to issues like bike clustering and supply-demand imbalances \citep{pan2019deep}. To address the challenge of maintaining supply and demand equilibrium in FFBSS, researchers have employed deep learning algorithms for predicting demand and supply dynamics \citep{kim2019graph, xu2020deep, mehdizadeh2022bike}. Additionally, optimization algorithms such as linear programming \citep{leclaire2018method}, integer programming \citep{raviv2013static}, genetic models \citep{gao2019moment}, and dynamic programming \citep{caggiani2018modeling} are used for efficient fleet rebalancing.

While prior studies have effectively used optimization algorithms, there are still limitations. Few models comprehensively incorporate multi-sourced data, and research on the congestion level of shared bicycle parking spots during specific periods, especially morning peak periods, remains limited. In this study, we aim to address these gaps by examining the congestion of shared bicycle parking spots during the morning peak period on Xiamen Island. This research provides valuable insights into user behavior and urban traffic patterns. The primary contributions of this study can be summarized as follows:

\begin{itemize}
    \item Exploration of Relationships: Investigating the intricate connections between parking spot congestion, urban traffic patterns, user behaviors, and Points of Interest (POI) data within the framework of Free-Floating Bike-Sharing Systems (FFBSS).

    \item Alignment Discovery: Highlighting the synergies among bike-sharing routes, established subway lines, and bus stops, revealing the prevalence of a "subway + bike-sharing" and "bus + bike-sharing" paradigm.
    
    \item Introduction of Congestion Metric: Pioneering the incorporation of a novel variable, congestion density ($C$), designed to quantify the dynamic levels of congestion in bicycle-sharing systems. This metric significantly contributes to the identification of congested bike-sharing (BS) parking spots.
    
    \item Comprehensive Clustering Framework: Developing a sophisticated framework for clustering congested spots that encompasses various factors, including congestion density ($C$), parking capacity ($P_c$), bike-sharing order counts, and Points of Interest (POI) data. This framework provides a holistic approach to understanding and managing congestion in shared bicycle systems.
\end{itemize}

The remainder of this paper is structured as follows: Section 2 reviews relevant research; Section 3 introduces the research area and data sources; Section 4 outlines the methodology; Section 5 presents the experimental results and discussions; and Section 6 summarizes the findings and outlines future directions.

\section{Related work}\label{sec:related-work}

This section provides an overview of related research on the analysis of shared bike data, understanding travel patterns and spatial clustering, managing and optimizing bike-sharing systems, and identifying crowded parking spots. These topics are crucial for enhancing travel efficiency, alleviating congestion, and promoting sustainable urban mobility.

    \subsection{Travel Patterns and Bike-sharing Data Analysis}

Shared bicycle systems have gained popularity as environmentally friendly and cost-effective transportation modes, addressing gaps in existing public transit networks. These systems offer a convenient alternative for commuters \citep{midgley2009role}. Recent research has focused on analyzing bike-sharing data to uncover travel patterns and spatial clustering resulting from cycling behavior.

For example, \cite{zhang2021mobility} utilized shared bicycle data from Singapore to reveal distinct travel patterns and spatial clustering related to various travel purposes. \cite{chang2020understanding} proposed a topic-based two-stage algorithm using station-free shared bike data to understand user travel behavior and discover functional regions within cities. \cite{xing2020exploring} studied Mobike bicycles and Point of Interest (POI) data to identify shared bicycle parking locations based on different travel purposes, providing insights into residents' daily activity patterns.

In addition to travel patterns and spatial clustering, several studies have explored other aspects of bike-sharing systems. \cite{xu2023exploring} investigated core population flow structures, \cite{saberi2018understanding} examined the impacts of transit disruptions on bike-sharing systems, and \cite{kager2016characterisation} analyzed the integration of bike-sharing systems with public transportation. These studies contribute to a comprehensive understanding of bike-sharing systems and their interactions with urban mobility.

Furthermore, the COVID-19 pandemic has had a significant impact on shared bicycle usage. \cite{teixeira2021motivations} emphasize the importance of BSS in resilient urban transport systems during health crises and advocate for continued BSS operation as a means to preserve physical distance and promote sustainable transportation alternatives. \cite{shi2023exploring} and \cite{hu2021examining} have modeled the spatiotemporal evolution of shared bicycle usage during the pandemic, providing valuable insights into mobility patterns and the effects of the crisis on shared transportation systems.

    \subsection{Bike-Sharing System Management and Optimization}

Efficient management and optimization play a crucial role in addressing the issue of supply-demand imbalance in shared bicycle systems. \cite{raviv2013static} provided valuable insights into the static repositioning of bike-sharing systems, offering models and solution approaches to mitigate congestion problems. \cite{caggiani2018modeling} presented a modeling framework for the dynamic management of free-floating bike-sharing systems, highlighting the significance of dynamic management in optimizing bike distribution and alleviating congestion at parking spaces.

To enhance decision-making and operational efficiency, researchers have explored advanced prediction and optimization techniques. \cite{kim2019graph} proposed a Graph Convolutional Network approach to predict hourly bike-sharing demands, incorporating spatial, temporal, and global effects. \cite{xu2020deep} introduced a deep learning-based multi-block hybrid model for bike-sharing supply-demand prediction. These methods improve the accuracy of demand forecasting and facilitate effective resource allocation in shared bicycle systems.

In addition to demand prediction, incentive mechanisms and congestion management strategies have been investigated. \cite{wang2021two} proposed a two-stage incentive system for free-floating bike sharing, incorporating pick-up and drop-off incentives based on user preferences. This approach aims to incentivize users to redistribute bicycles more evenly, addressing the issue of imbalanced bike distribution. \cite{mehdizadeh2022bike} employed deep learning models to predict the spatiotemporal demand characteristics of shared bicycles, providing a better understanding of demand patterns and supporting operational management.

Furthermore, researchers have focused on understanding spatial heterogeneity and optimizing parking locations and capacities. \cite{he2022geographically} specifically studied the bike-sharing system in Nanjing, China, utilizing geographically weighted multinomial logit models to comprehend spatial heterogeneity in bike-sharing renting-returning imbalances. This research offers insights into congestion management and resource allocation tailored to specific geographical areas. \cite{zhao2021geo} constructed a hybrid clustering framework for identifying shared bicycle parking locations and capacities, achieving more effective coverage compared to current practices by optimizing parameter selection.
    
    \subsection{Identifying Crowded Parking Spots for Shared Bicycles}

The identification of crowded parking spots for shared bicycles is of utmost importance for addressing traffic congestion, ensuring an adequate supply of bicycles for users, and facilitating efficient repositioning by operators. \cite{wang2016applying} employed spatial-temporal analysis and retail location theory to select public bike sites in Taipei, shedding light on the factors influencing the success of bike-sharing programs. \cite{keler2020extracting} analyzed trip destination data to extract destination hotspots specific to commuters, comparing taxi services with Citi Bike in NYC and providing insights for strategic station placement.

Expanding on existing research, \cite{hui2022hotspots} identified hotspots for bicycle return and pickup in Xiamen and developed an evaluation framework based on demand-supply imbalances, land use, and demand characteristics. They categorized hotspots into overloaded areas, areas requiring improved service quality, and stable zones.\cite{choi2023combatting} conducted hotspot analysis on shared bikes and utilized Shapley Additive Explanations (SHAP) values to understand the impact of various factors on mismatches, identifying both hot and cold spots.

Building upon the aforementioned studies, our research offers a comprehensive depiction of urban mobility by integrating data from dockless shared bicycles, subway routes, bus stations, and Points of Interest (POI). Our innovative framework aims to identify and classify hotspots for shared bicycles. The outcomes of our study make significant contributions to the optimization of bike-sharing systems, congestion mitigation, and the promotion of sustainable urban mobility.

\section{Study Area and Datasets}\label{sec:3}

This study focuses on the dockless bike-sharing system located on Xiamen Island, China. Xiamen is a dynamic city in the southeast of Fujian Province, known for its historical charm and modern vitality. The bike-sharing initiative was introduced in December 2016 and quickly became a popular mode of transportation for short distances within the city. In July 2020, the Xiamen Municipal Law Enforcement Bureau implemented measures to improve the bike-sharing system, including the adoption of high-precision split locks and electric fences based on Beidou positioning technology. These improvements significantly enhanced parking efficiency and addressed the issue of indiscriminate parking of shared bicycles. Currently, there are 15,202 electronic fence parking spaces within the shared bicycle strict control area. According to the "2020 Xiamen Urban Transportation Development Report," shared travel modes, particularly online ride-hailing and shared bicycles, experienced substantial growth in 2020. The total number of shared bicycle rides reached 113.79 million, a remarkable 30

The utilization of shared bicycles in Xiamen is concentrated in the Siming and Huli Districts, which are the city's economic, political, cultural, and financial hub. Therefore, this study specifically focuses on these two districts, which have a combined population exceeding 2 million according to the Xiamen Statistical Yearbook, 2021. The travel patterns of Xiamen residents reveal two distinct peaks in a day, with the morning peak showing a higher concentration of shared bicycle usage. Consequently, this research focuses on the morning peak hours to examine the challenges related to shared bicycle parking congestion, alleviate traffic pressures, and improve overall travel efficiency.

\subsection{Dockless Bike-Sharing Data}
This study utilizes a comprehensive dockless bike-sharing dataset obtained from the official Digital China Innovation Contest (DCIC 2021) websites. The dataset includes bike-sharing order information, trajectory data, and electric fence data. It provides detailed records of key elements such as bicycle ID, fence ID, latitude, longitude, lock status, and update time. The dataset covers 198,382 journeys during the morning peak hours (6:00–10:00) across five weekdays in December 2020, offering insights into the spatiotemporal dynamics of shared bicycle mobility. Despite the winter season, Xiamen's average maximum temperature of around 20°C makes cycling an attractive option. The dataset includes 14,071 specified shared bicycle parking spots within the study area, with 594 remaining idle during the survey period, indicating no orders were generated. This granular dataset focuses on morning peak hours and reveals the utilization patterns of shared bicycle parking spaces, providing profound insights into the dynamics of urban mobility in Xiamen.

\subsection{Point of Interest (POI) Data}
The POI data used in this study is sourced from Amap (\url{https://www.amap.com}). It plays a significant role in urban land use classification and travel pattern analysis. The data consists of approximately 5,999 POI data points centered on Xiamen Island. Each data point includes attributes such as name, category, geographical coordinates (longitude and latitude), and geographic region. The POI data is categorized into nine distinct categories: 'Transport,' 'Shopping,' 'Culture,' 'Sports,' 'Life,' 'Landscape,' 'Restaurant,' 'Medical,' and 'Company.' This categorization helps explore the daily travel patterns and trip purposes of dockless bike-sharing users.

\subsection{Subway Line Data and Bus Station Data}
The integration of subway line data and bus station information, sourced from Amap, is crucial for understanding urban activity patterns and the dynamic synergy between public transportation systems. This dataset improves mobility planning by identifying locations near bus and subway stations that experience a high concentration of bike-sharing activity. By extracting the latitude and longitude of subway stations, bus stops, and lines, valuable insights into the integrated use of subway, bus, and bike-sharing systems by residents are obtained, promoting sustainable urban mobility practices. This approach provides a nuanced understanding of the interplay between public transportation systems and bike-sharing behaviors, offering actionable insights for optimizing transportation networks and enhancing overall efficiency in urban mobility in Xiamen.

\begin{figure}[ht]
            \begin{center}
            \includegraphics[width=0.8\linewidth]{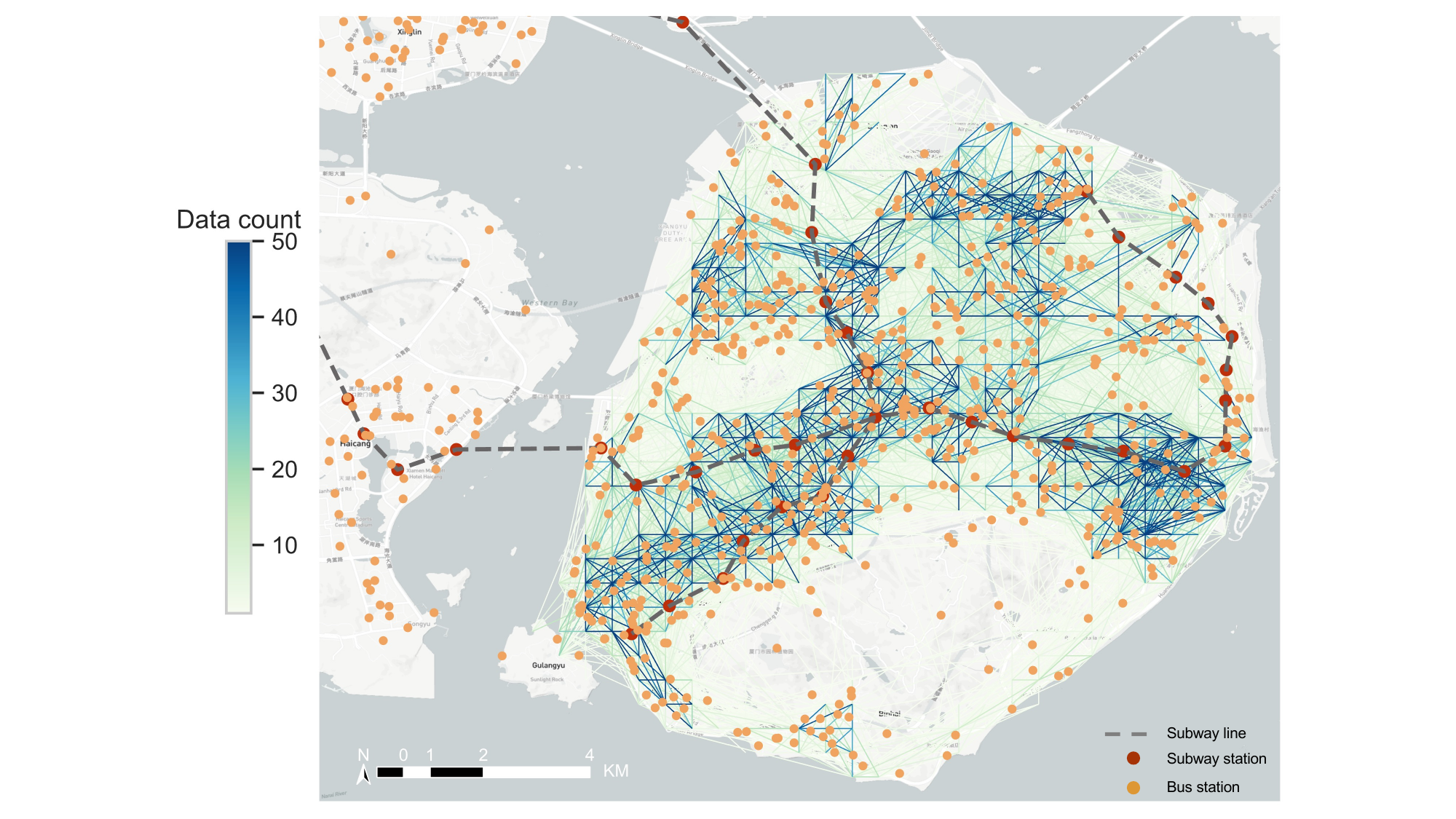}
            \end{center}
            \caption{bike-sharing OD (Origin-Destination) trajectories map}\label{fig:OD_map}
        \end{figure}

Fig.~\ref{fig:OD_map} visually illustrates the distribution of bike-sharing Origin-Destination (OD) trajectories, with shading indicating the intensity of bike travel between different origin and destination pairs. Notably, bike-sharing orders are more prevalent near subway routes and bus stations compared to other regions, highlighting the strategic influence of public transportation nodes on shared bicycle usage patterns.

To conduct our analysis, we integrate data from diverse sources, including dockless bike-sharing, subway lines, bus stations, and points of interest. We leverage information from Amap, OpenStreetMap, and Mapbox to ensure a meticulous understanding and categorization of popular bike-sharing locations within the designated study area. This strategic integration of various data sources serves as the foundation for our comprehensive exploration of the intricate dynamics of shared bicycle mobility on Xiamen Island. Table~\ref{tab:data_attributes} provides an excerpt of the data extracted from these multiple sources.

\begin{table}[ht]
    \caption{Description of Data Samples}
    \label{tab:data_attributes}
    \centering
    \fontfamily{ptm}\selectfont 
    \begin{tabular}{lp{3.5cm}p{6cm}}
        \toprule
        \textbf{Data Attribute} & \textbf{Type} & \textbf{Example} \\
        \midrule
        Bicycle\_ID & String & 00015277***********8e8902703 \\
        Update\_time & DateTime & 2020-12-21 08:17:12 \\
        Bike\_coordinates & List of Floats & [118.140385, 24.521156] \\
        Lock\_status & Integer & 0 (unlocked) or 1 (locked) \\
        Fence\_ID & String & Changle Road 0\_L\_A17001 \\
        Fence\_LOC & List of Lists & [[118.103198,24.527338], [118.103224,24.527373], [118.103236,24.527366], [118.103209,24.527331], [118.103198,24.527338]] \\
        Subway\_station & String & Wusshipu \\
        Subway\_line & String & Line 1 \\
        Subway\_station\_coordinates & List of Floats & [118.121573, 24.501580] \\
        Bus\_stop & String & BRT Shuangshi Middle School Site \\
        Bus\_stop\_coordinates & List of Floats & [118.150136, 24.519946] \\
        POI\_name & String & Zhongshan Park \\
        POI\_address & String & Gongyuan South Road (next to Exit 3A of Zhongshan Park Metro Station) \\
        POI\_district & String & Siming District \\
        POI\_type & String & Landscape \\
        POI\_coordinates & List of Floats & [118.090086, 24.459141] \\
        \bottomrule
    \end{tabular}
\end{table}

\section{Methodology}\label{sec:methodology}

In this section, we present our methodology for analyzing urban daily activity patterns and identifying and classifying congested bike-sharing (BS) parking spots.

\subsection{Methodology Framework}

\begin{figure}[ht]
    \centering
    \includegraphics[width=0.8\linewidth]{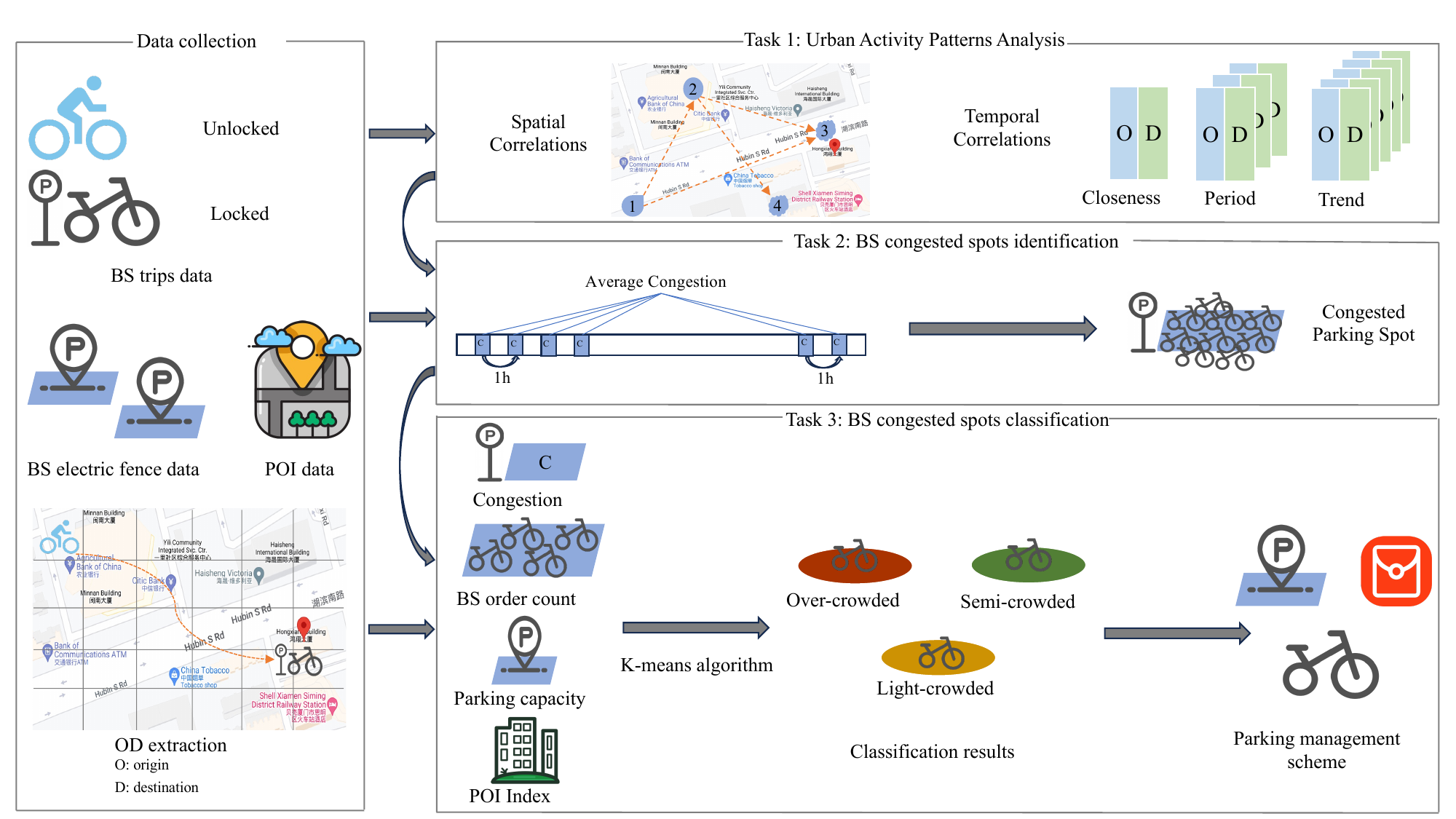}
    \caption{Model Framework}
    \label{fig:Model_Framework}
\end{figure}

As illustrated in Fig.~\ref{fig:Model_Framework}, our approach consists of two layers: the data layer, involving data collection from various sources, and the model layer, which addresses three core research tasks.

\subsection{Urban Mobility Patterns Analysis}

Understanding urban mobility patterns is crucial for effective city planning, transportation management, and enhancing overall urban livability. To achieve this, we refined our dataset by filtering out irrelevant data points, such as bike journeys exceeding 10 kilometers or less than 100 meters. We then focused on extracting Origin-Destination (OD) trajectory data to unveil the spatial attributes of the bike-sharing ecosystem on Xiamen Island.

Additionally, we analyzed subway line and bus data to uncover the intricate relationship between public transportation networks and bike-sharing activities. Points of Interest (POI) data were incorporated to explore human mobility patterns, elucidating the connection between bike-sharing hubs and various POI categories. Our approach involved integrating diverse data streams and employing spatial and temporal analytical methodologies, providing valuable insights with implications for urban planning and mobility management strategies.

\subsection{Identification of Congested Bike-Sharing (BS) Parking Spots}

This study aims to identify and quantify congestion levels at shared bicycle (BS) parking spots during the morning peak hours on Xiamen Island, offering crucial insights for optimizing urban transportation planning.

To quantify congestion levels, a dedicated formula was developed:

\begin{equation}\label{eq_C}
    C_{i,j} = \frac{P_{i,j}}{Pc_i} = \frac{Inflow_{i,j} - Outflow_{i,j}}{Pc_i}
\end{equation}

In this formula, $C_{i,j}$ represents congestion density at parking spot \(i\) during time interval \(j\). Here, \(P_{i,j}\) is the real-time number of shared bicycles parked at parking spot \(i\) during time interval \(j\). The parameters \(Inflow_{i,j}\) and \(Outflow_{i,j}\) denote the total inflow and outflow during the specific time interval. The value of \(P_c\), referred to as \textit{Parking Capacity}, signifies the maximum number of bicycles a parking spot can accommodate. Following the Ministry of House and Urban-Rural Development's guidelines, a single parking space should ideally measure 0.6-0.8$m$ in width and 2.0$m$ in length, corresponding to an area of 1.2-1.6 $m^2$. Typically, the minimum value of 1.2 $m^2$ is adopted as the standard for a single parking spot to optimize space usage and reduce operational costs.

As \(P_{i,j}\) approaches the total parking capacity, \(C_{i,j}\) tends towards 1, indicating relative congestion. Conversely, when \(P_{i,j}\) significantly lags behind the total capacity, \(C_{i,j}\) approaches 0, suggesting relative emptiness. Moreover, the formula considers the net effect of inflow and outflow, providing valuable insights into shared bicycle movement dynamics within specific time intervals. Overall, it offers a quantitative perspective on congestion levels at shared bicycle parking spots, contributing to a comprehensive understanding of utilization patterns within the bike-sharing system during morning peak hours on Xiamen Island.

\subsection{Classification of Congested Bike-Sharing (BS) Parking Spots}

To classify congested bike-sharing (BS) parking spots, this study integrates four key factors: congestion density (\(C\)), parking capacity (\(P_c\)), bike-sharing order counts, and Points of Interest (POI) data. The objective is to explore the spatial distribution and characteristics of congested parking spots during morning peak hours on Xiamen Island.

\begin{itemize}
    \item \textbf{Congestion Density (\(C\)):} The previously defined \(C_{i,j}\) values will serve as the primary measure of congestion, capturing the real-time dynamics of shared bicycle parking spots.
    
    \item \textbf{Parking Capacity (\(P_c\)):} The maximum capacity of each parking spot is crucial in understanding how close the spot is to its full potential during peak hours.
    
    \item \textbf{Bike-Sharing Order Counts:} The order counts at each parking spot provide additional insights into user demand and the popularity of specific locations.
    
    \item \textbf{Points of Interest (POI) Data:} To gauge the influence of the surrounding environment on parking spot congestion, a POI index (\(PI\)) is calculated. This index represents the proportion of aggregated counts in four specified POI categories concerning the total number of POI data points. The formula is expressed as:
    
    \begin{equation}\label{eq_PI}
        PI_{i} = \frac{number \,of \,POIs(i,j)}{number \,of \,POIs}
    \end{equation}
    
    where \(number \,of \,POIs(i,j)\) signifies the number of POIs at parking spots \(i\) belonging to category \(j\) within a maximum walking radius of 300m. \(number \,of \,POIs\) denotes the aggregate number of POIs on Xiamen Island.
\end{itemize}

Inspired by traffic classification in Yao et al.'s work\citep{yao2020encrypted} and BBS classification in Godichon et al.'s work\citep{godichon2019clustering}, our approach involves the application of the Gaussian Mixture Model (GMM) and K-means clustering algorithms for the classification of urban bike-sharing congested parking spots. These methods group parking spots based on the multidimensional feature space defined by the chosen factors.

\subsubsection{Gaussian Mixture Model (GMM) Clustering}

The Gaussian Mixture Model (GMM) is a robust probabilistic model renowned for its ability to perform soft clustering \citep{reynolds2009gaussian}. It demonstrates flexibility in capturing complex data structures characterized by varying sizes, interrelationships, and subpopulation distributions.

At its essence, the GMM employs an objective function to quantify the likelihood of data given the mixture model. Mathematically, the GMM objective function is expressed as:

\begin{equation}
\text{GMM Objective Function: } L(\Theta) = \sum_{i=1}^{N}\log \left( \sum_{k=1}^{K} \pi_k \mathcal{N}(\mathbf{x}_i | \mathbf{\mu}_k, \mathbf{\Sigma}_k) \right)
\end{equation}

Here, \(N\) represents the number of data points, \(K\) is the number of Gaussian components in the mixture, \(\pi_k\) is the mixing weight of the \(k\)-th component (ensuring \(\sum_{k=1}^{K} \pi_k = 1\)), \(\mathbf{x}_i\) is a data point, \(\mathbf{\mu}_k\) is the mean of the \(k\)-th Gaussian, \(\mathbf{\Sigma}_k\) represents the covariance matrix of the \(k\)-th Gaussian, and \(\mathcal{N}\) is the probability density function of a Gaussian.

The parameters of the GMM (\(\Theta = \{\pi_k, \mathbf{\mu}_k, \mathbf{\Sigma}_k\}\) for all \(k\)) are typically estimated using the Expectation-Maximization (EM) algorithm. EM iteratively performs the Expectation (E) step, estimating posterior probabilities, and the Maximization (M) step, updating parameters to maximize likelihood based on these probabilities.

While GMM stands out as a powerful clustering algorithm capable of capturing intricate relationships in data through probabilistic modeling, it's crucial to acknowledge its computational expense for large datasets due to the iterative nature of the EM algorithm.

\subsubsection{K-means Clustering}

K-means clustering, a widely adopted unsupervised learning algorithm, is celebrated for its simplicity and efficiency in partitioning datasets into distinct clusters\citep{macqueen1967some}. The algorithm operates on a straightforward premise: assigning each data point to one of \(K\) clusters while minimizing within-cluster variances and maximizing between-cluster distances.

The performance of K-means is evaluated through its objective function, commonly known as inertia or within-cluster sum of squares (WCSS). Mathematically, the K-means objective function is expressed as:

\begin{equation}
\text{K-means Objective Function: } J = \sum_{i=1}^{N} \sum_{k=1}^{K} r_{ik} |\mathbf{x}_i - \mathbf{\mu}_k|^2
\end{equation}

Here, \(N\) represents the total number of data points, \(K\) is the specified number of clusters, \(r_{ik}\) is a binary indicator function (1 if data point \(\mathbf{x}_i\) is assigned to cluster \(k\), 0 otherwise), \(\mathbf{\mu}_k\) represents the centroid of cluster \(k\), and \(\|\cdot\|\) denotes the Euclidean distance.

K-means iteratively minimizes the objective function through two steps: the assignment step (assigning each data point to the cluster with the closest centroid) and the update step (recalculating centroids as the mean of data points assigned to each cluster). It is widely used in various fields, including image segmentation, customer segmentation, anomaly detection, and recommendation systems. It provides a simple and interpretable way to group similar data points together based on their features or distance relationships. 

\section{Results and Discussions}\label{sec:Results}

In this section, we present our experimental findings and conduct a comprehensive analysis of the spatio-temporal mobility patterns of bike-sharing. As outlined in the Introduction, our research objectives are focused on understanding urban mobility behaviors, identifying congested bike-sharing (BS) parking spots, and categorizing these spots to improve bike-sharing parking management strategies.

\subsection{Urban Mobility Patterns Analysis}

\begin{figure}[ht]
    \centering
    \begin{subfigure}{0.45\textwidth}
        \centering
        \includegraphics[width=\linewidth]{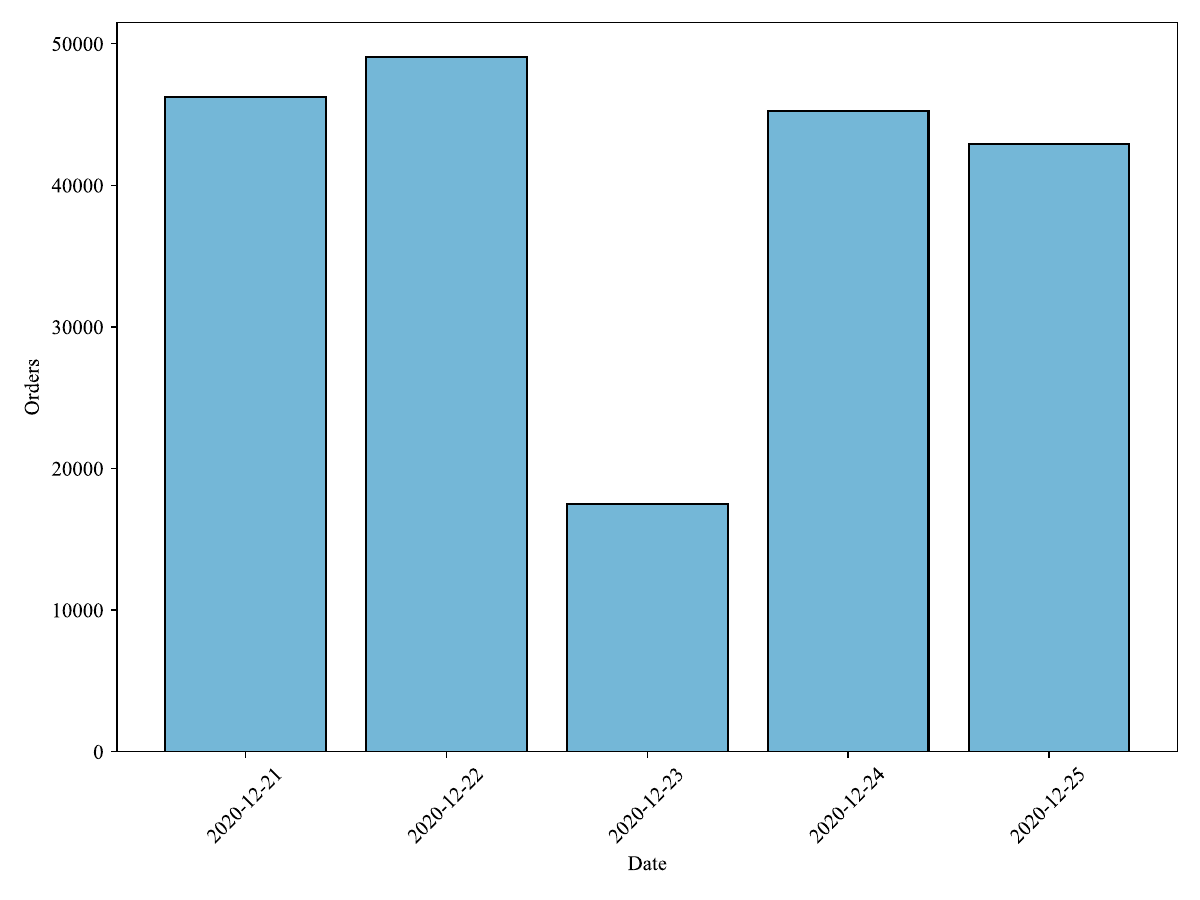}
        \caption{Daily distribution.}
    \end{subfigure}
    \hfill
    \begin{subfigure}{0.45\textwidth}
        \centering
        \includegraphics[width=\linewidth]{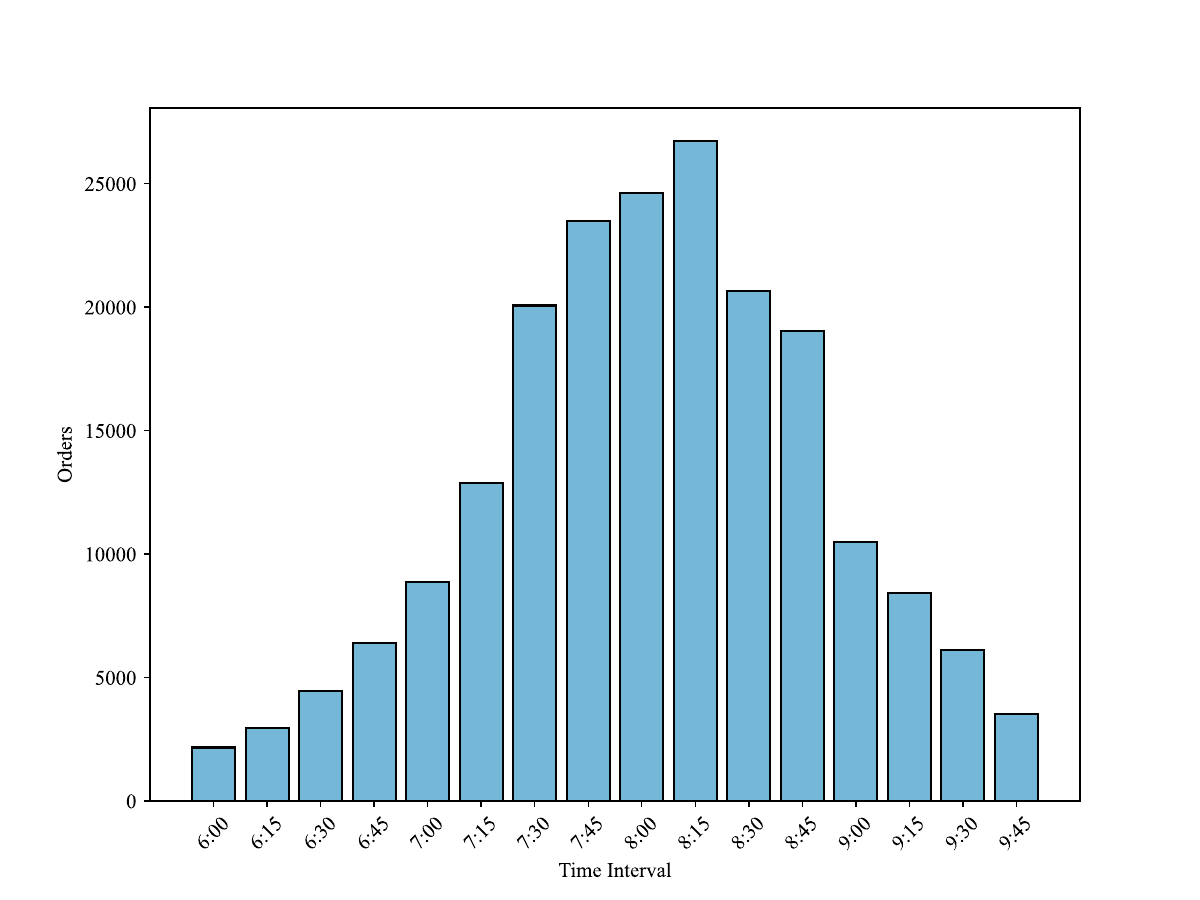}
        \caption{Temporal distribution.}
    \end{subfigure}
    
    \begin{subfigure}{0.45\textwidth}
        \centering
        \includegraphics[width=\linewidth]{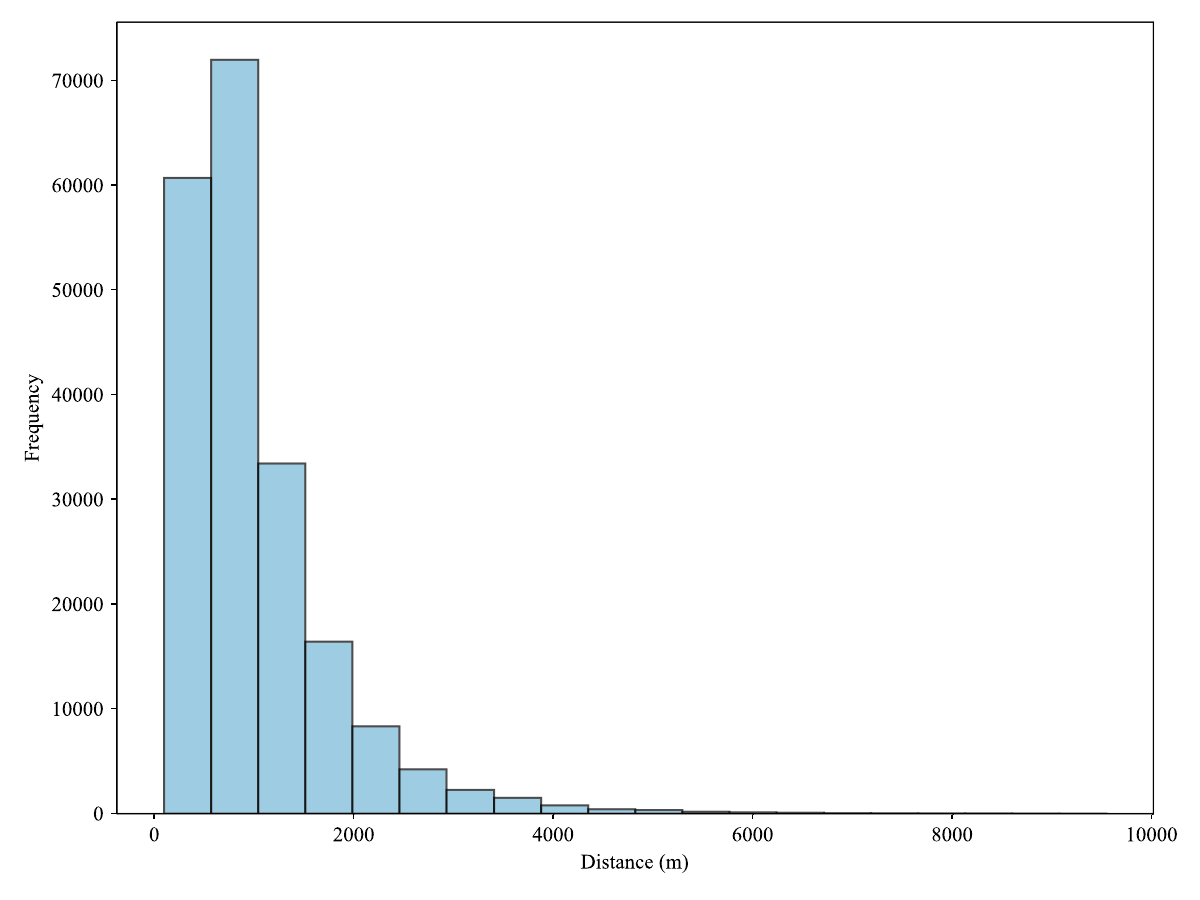}
        \caption{Travel distance distribution.}
    \end{subfigure}
    \hfill
    \begin{subfigure}{0.45\textwidth}
        \centering
        \includegraphics[width=\linewidth]{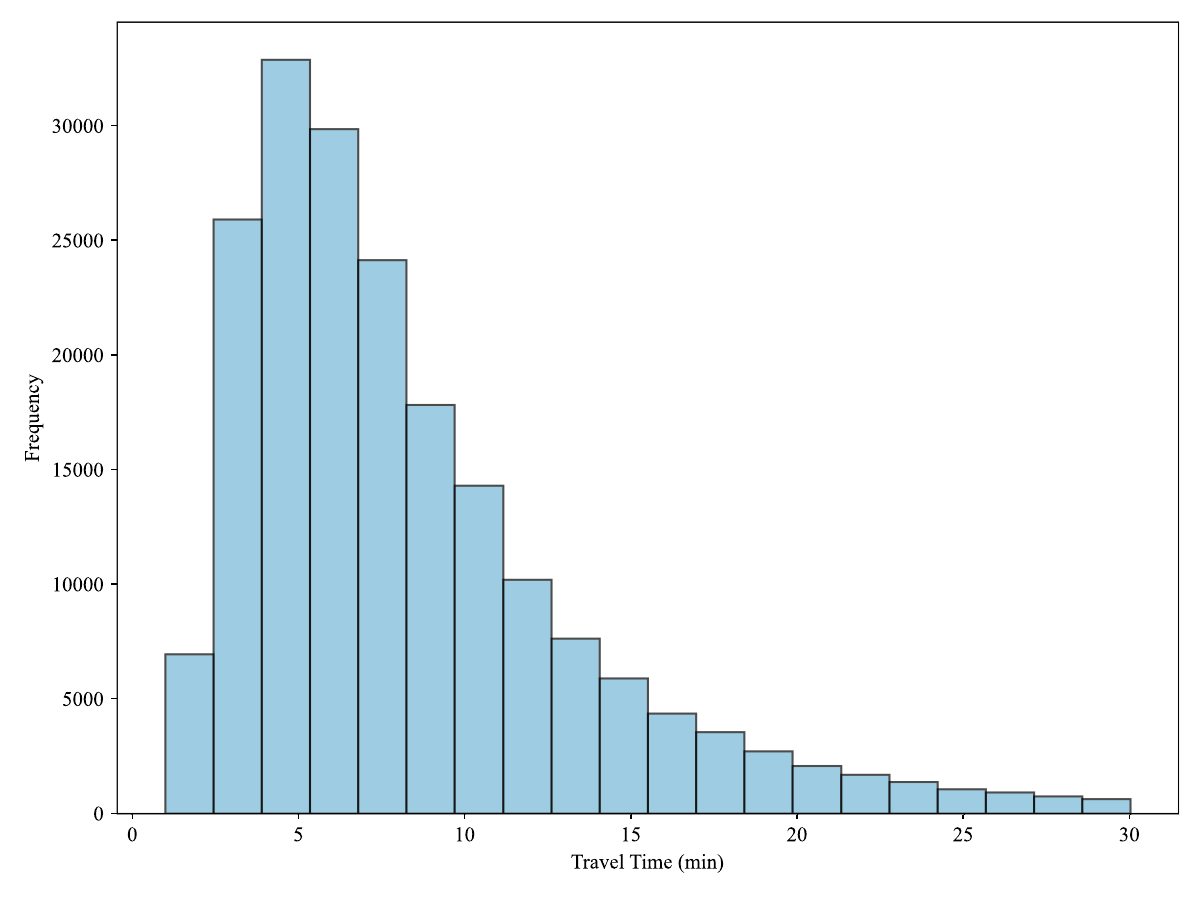}
        \caption{Travel time distribution.}
    \end{subfigure}
    \caption{Descriptive analysis of BSS. (a) Daily distribution. (b) Temporal distribution. (c) Travel distance distribution. (d) Travel time distribution.}
    \label{fig:descriptive_analysis}
\end{figure}

The descriptive analysis is visually presented in Fig.~\ref{fig:descriptive_analysis}. The bar chart in Fig.~\ref{fig:descriptive_analysis}(a) depicts the daily order counts for five days, showing a significant decrease on December 23rd due to moderate rainfall, making cycling less attractive. Subsequent analysis is based on OD data from the remaining four weekdays. Fig. \ref{fig:descriptive_analysis}(b) shows that the highest order volume occurred between 08:00 and 08:15. Investigating transportation patterns, Fig. \ref{fig:descriptive_analysis}(c) reveals that approximately 70\% of travel distances using the BS mode are within 1,000 meters, underscoring the importance of BS as a primary mode for covering the initial or final leg of journeys. Additionally, Fig. \ref{fig:descriptive_analysis}(d) illustrates the distribution of trip durations, with most trips taking around 10 minutes, though some extend beyond 20 minutes.

\begin{figure}[ht]
    \begin{center}
        \includegraphics[width=0.8\linewidth]{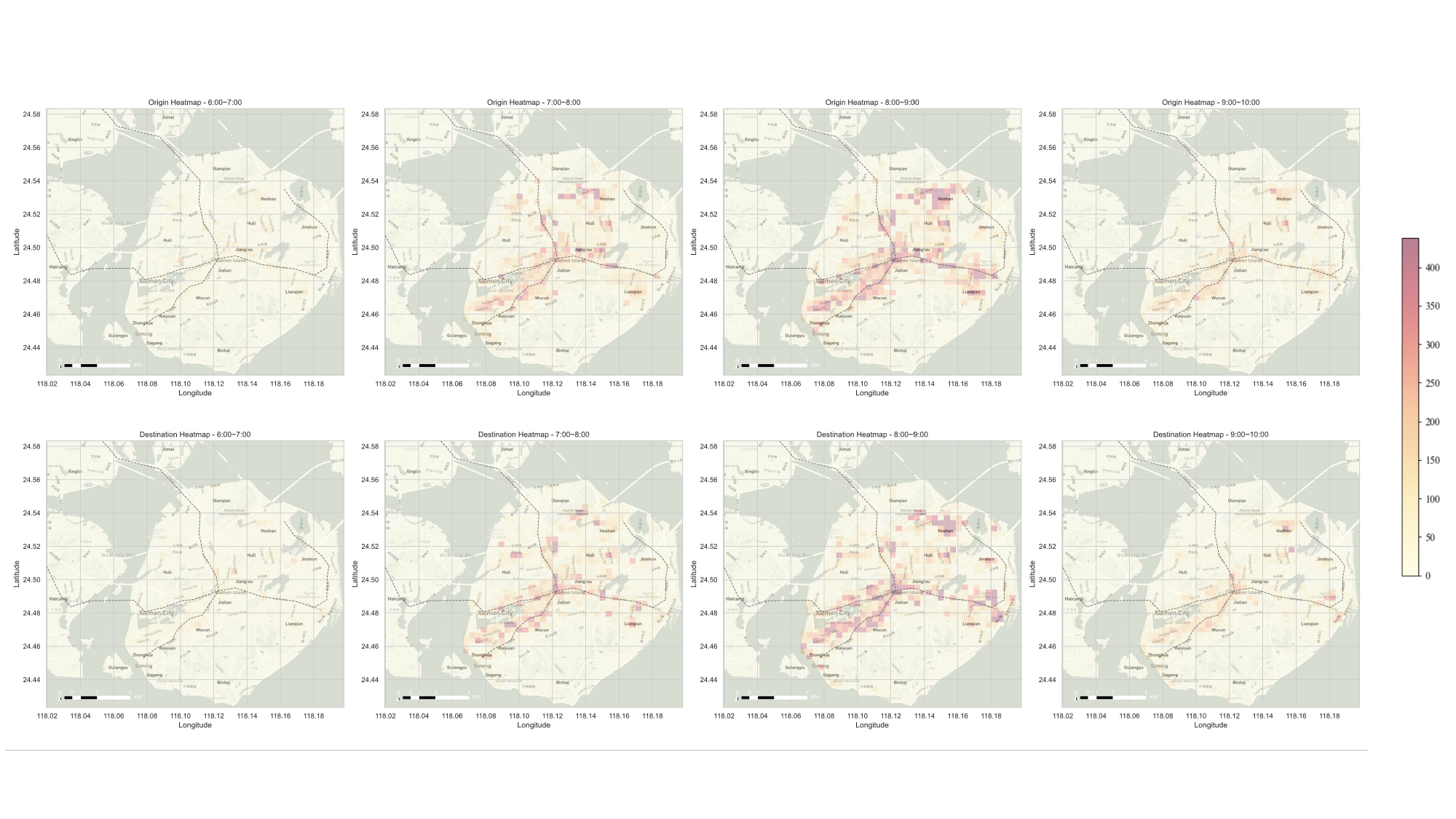}
    \end{center}
    \caption{Spatial distribution of bike-sharing journeys during different periods in the morning peak. From top to bottom are the origin and destination maps.}
    \label{fig:combined_heatmaps}
\end{figure}

Fig.~\ref{fig:combined_heatmaps} illustrates the geographical distribution of BS travels during the morning peak across four days. High-density travel regions near subway lines from 8:00 to 9:00 emerge as focal points where commuters predominantly utilize BS, emphasizing its role in facilitating efficient last-mile commuting, promoting subway integration, and highlighting the necessity for resource allocation during peak hours.

\begin{figure}[ht]
    \centering         
    \begin{subfigure}{0.45\textwidth}
        \centering
        \includegraphics[width=\linewidth]{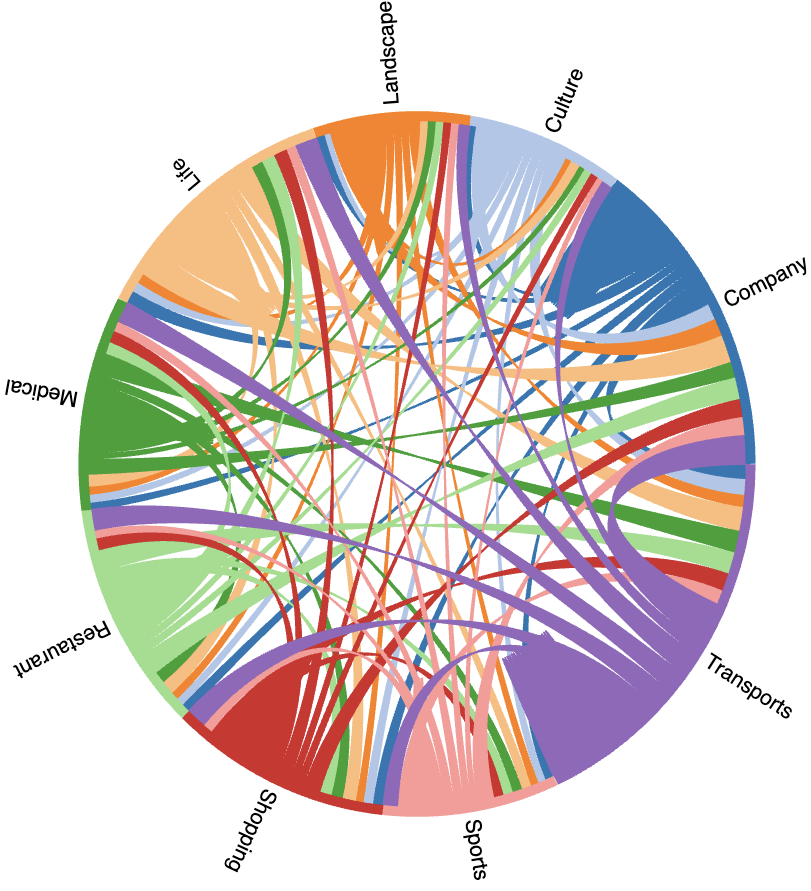}
        \caption*{(a)}
    \end{subfigure}
    \hfill
    \begin{subfigure}{0.45\textwidth}
        \centering
        \includegraphics[width=\linewidth]{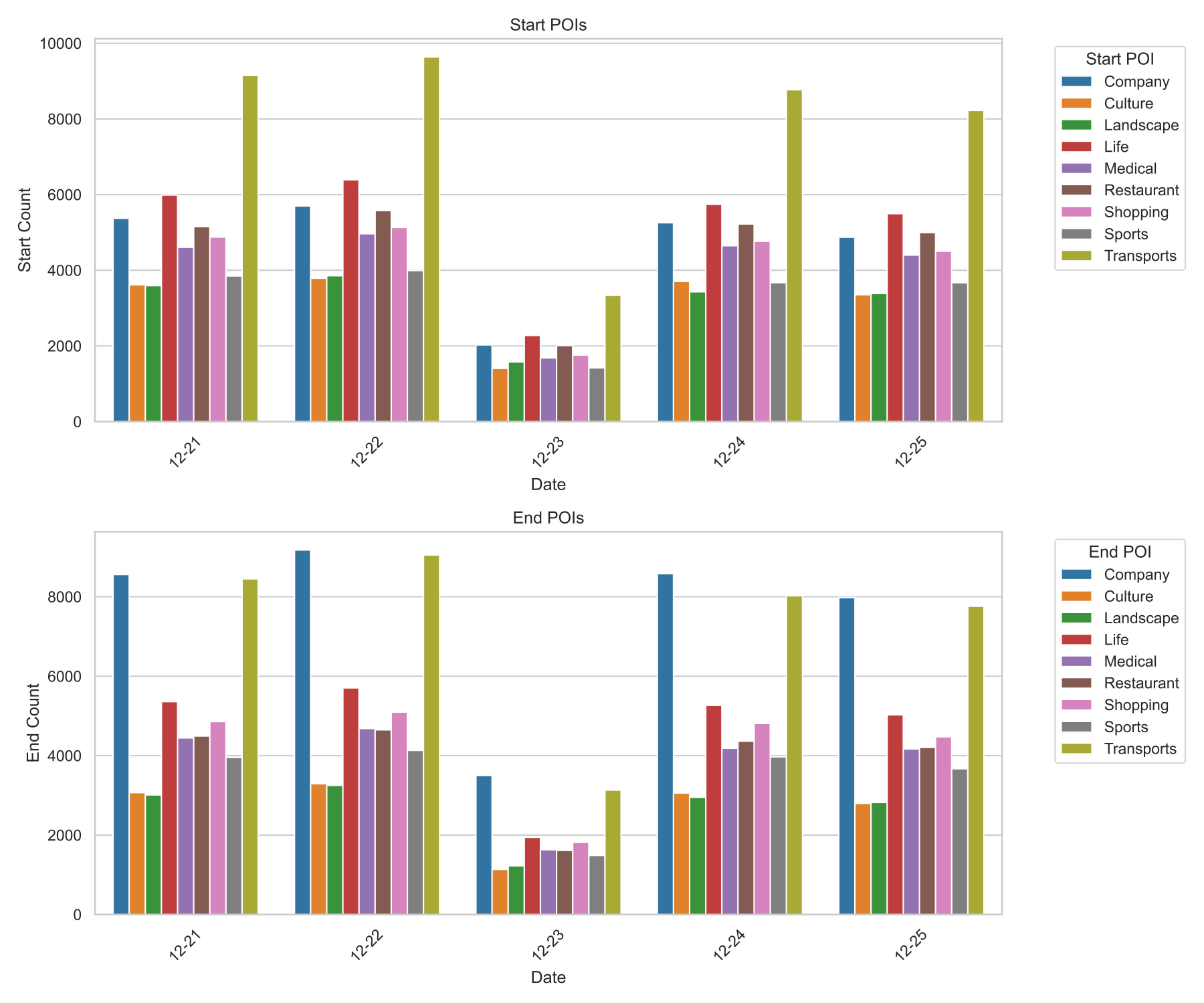}
        \caption*{(b)}
    \end{subfigure}
    \caption{Trip patterns of BS based on POI data. (a) A chord diagram of the mobility of bike-sharing between different POI types. Nodes in distinct colors represent different activity types, with node lengths denoting their respective proportions. The width of the central connecting lines reflects the volume of transfers between these activities. (b) Distribution of bike-sharing journeys based on starting and ending POIs.}
    \label{fig:POI}
\end{figure}

The trip patterns of dockless bike-sharing systems are influenced by various factors such as dining, transportation, shopping, work, and residential places \citep{xing2020exploring}. To categorize points of interest (POI) data and unveil daily activity patterns and travel purposes, nine categories were identified, including Transports, Shopping, Culture, Sports, Life, Landscape, Restaurant, Medical, and Company. Fig.~\ref{fig:POI}(a) illustrates transitions between these activities, highlighting BS's heavy utilization in key POIs like Transports (29.24\%), Company (20.04\%), and Life (16.92\%). Notably, Transportation emerges as the primary category, emphasizing BS's vital role in connecting transportation hubs to various activities. Fig.~\ref{fig:POI}(b) showcases BS journeys, indicating that transportation hubs are popular starting points, while "Company" locations are common return points, emphasizing BS's convenience for short-distance commutes. The "Transports" category is prominent for both starting and ending points, emphasizing its popularity for daily commutes, while moderate usage in "Culture," "Medical," and "Sports" suggests recreational use. Furthermore, notable usage in the "Shopping" and "Restaurant" categories showcases BS's facilitation of shopping and dining experiences, while balanced use in "Life" and "Landscape" reflects diverse BS usage across activities.

\subsection{Results: Identification of Congested Shared Bicycle Parking Spots}

This study focuses on identifying congested shared bicycle parking spots during the early morning peak hours. By systematically correlating latitude and longitude with electronic fence IDs, we computed pick-up and return counts in hourly intervals. The congestion level for each parking spot was determined using Equation (\ref{eq_C}) outlined in the methodology. Congested spots were defined where the calculated congestion level exceeded a threshold of 1, signifying instances where demand surpasses parking capacity (\(Pc_i\)). This approach provides valuable insights into areas with heightened shared bicycle usage, aiding in urban mobility planning during the morning rush hour.

\begin{figure}[ht]
    \begin{center}
        \includegraphics[width=0.8\linewidth]{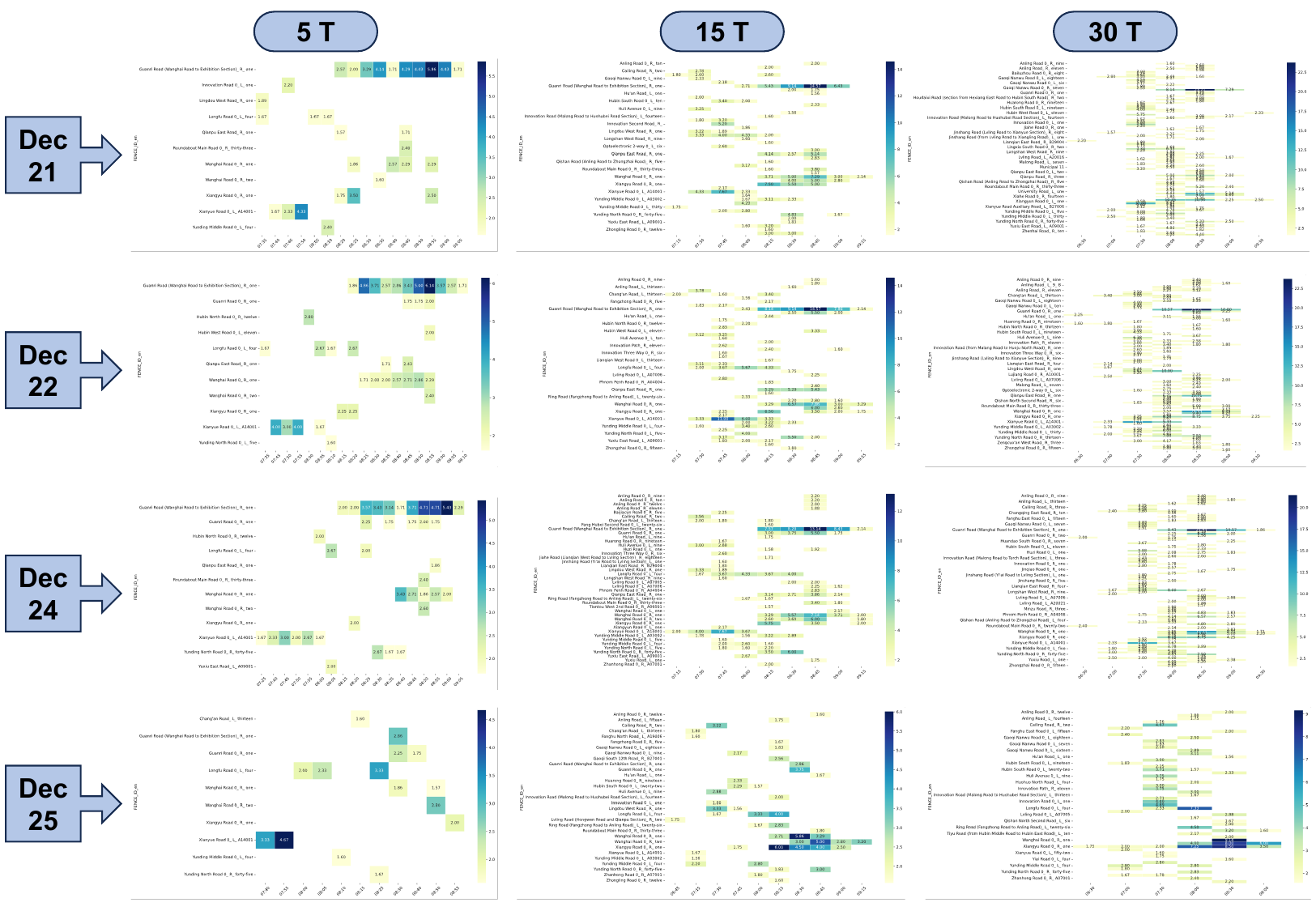}
    \end{center}
    \caption{Result of identifying congested BS parking spots at different time intervals. For clearer visuals, we display only instances where congestion exceeds 1.5, indicating shared bicycles parked exceed 1.5 times the total parking capacity. The color intensity in each cell corresponds to the congestion level, with darker shades indicating higher congestion.}
    \label{fig:congestion_spots}
\end{figure}

To capture the dynamic nature of congestion, Equation (\ref{eq_C}) was applied at various intervals, including 5, 15, and 30 minutes. The heatmap in Fig.~\ref{fig:congestion_spots} illustrates congestion patterns at different times during morning rush hours. Since the investigations occurred on weekdays, a certain spatiotemporal correlation in shared bicycle parking space congestion emerged. Severely congested spots were concentrated in areas like Guanri Road, Xianyue Road, Xiangyu Road, Wanghai Road, and Yunding North Road, which host numerous companies. Among them, Guanri Road (Wanhai Road to Hui Zhan Road\_R\_1) exhibited the most severe congestion, reaching a level of 6.14 at 08:55:00 on December 22 within a 5-minute interval, indicating severe overcrowding. Moreover, the period from 8:00 to 8:30 emerged as the most congested in the morning, aligning with the peak commuting period.

Subsequently, we selected data for parking spots in a congested state (\(C > 1\)), grouping it by electric fences and calculating the average congestion level for each spot using Equation (\ref{eq:Average Congestion}).

\begin{equation}\label{eq:Average Congestion}
    \text{Average Congestion}: \overline{C} = \frac{\sum_{k=1}^{n} \text{C}(k)}{n}
\end{equation}

Here, \(n\) represents the number of data points in the observation period, and \(\text{C}(k)\) is the congestion level for the \(k\)-th parking space. The inherent flexibility of the dockless shared bicycle system, relying on dynamic changes in user pick-up and drop-off, allows many locations to maintain the supply-demand balance in a short time. For instance, "Yunding Middle Road 0\_L\_A03002" has an average congestion level of 3.46 in 30-minute intervals but only 0.19 in 5-minute intervals. Despite equilibrium in some areas, our analysis revealed 563 congested shared bicycle parking spots.

\subsection{Results of Congested BS Parking Spots Classification}

To comprehensively analyze the spatial distribution and characteristics of congested parking spots and urban mobility patterns, we employed both the Gaussian Mixture Model (GMM) and K-means, considering parameters such as congestion levels (\(C\)), parking capacity (\(P_c\)), bike-sharing order counts, and Points of Interest (POI) data for spot classification. After evaluating both models, we chose the K-means clustering approach based on the Silhouette Score, which resulted in a score of 0.637. Moreover, the optimal \(K\) value for K-means was determined using the Elbow Method, leading to the selection of \(K=3\). The resulting clusters are categorized as Over-crowded, Semi-crowded, and Light-crowded.

\begin{itemize}
    \item \textbf{Over-crowded:} Accounting for 3.02\% (17 out of 563) of congested spots, these sites exhibit a substantial surge in shared bicycle activity that far surpasses the designated parking capacity. Despite possessing a relatively extensive parking capacity, it falls short of meeting the escalating demand for bike-sharing, underscoring a significant disparity between parking demand and supply in these particular areas. In terms of land utilization, this category boasts a higher number of Points of Interest (POI) compared to the other two, indicative of a more developed environment. Simultaneously, the prevalence of recreation and commercial POIs underscores a heightened demand for last-mile commuting to workplaces. Consider Xiamen Software Park, home to several software companies, whose employees mostly utilize bike sharing as a means of transportation to reach the park. Therefore, quick intervention measures are necessary, such as proactive bike dispatching and encouraging users to park next to relatively empty spots. To avoid service interruptions and guarantee that consumers have a smooth bike-sharing procedure during peak hours, prompt and immediate action is also required.
    
    \item \textbf{Semi-crowded:} Encompassing 15.28\% (86 out of 563) of congested spots, areas in this category maintain a moderate level of congestion, signaling that demand still surpasses parking capacity. Notably, locations such as Wushipu Metro Station and Lianban Metro Station, situated in Xiamen's bustling commercial hub, are surrounded by numerous corporate offices, shopping malls, and amenities. This strategic positioning has led the service provider to install several shared bicycle parking stations in this vicinity. During morning peak hours, these areas become focal points for commuters, resulting in increased demand for bike-sharing services. This type of spot is particularly prone to an insufficient supply of bike-sharing electric fences. In essence, allocating more or larger electric fences to the location could significantly alleviate parking congestion. Alternatively, using a red envelope incentive mechanism to encourage users to park at nearby spots can better align with demand based on the existing parking capacity.
    
    \item \textbf{Light-crowded:} Comprising 86.7\% (460 out of 563) of congested spots, locations in this category exhibit a relatively low level of congestion, indicating a general equilibrium between demand and capacity. Taking Xiamen Railway Station as an illustration, despite the high pedestrian traffic volume, individuals in this area primarily engage in brief activities. The equilibrium between supply and demand for bicycle-sharing parking spots remains stable, given the balanced arrival and departure of people, preventing abrupt fluctuations in parking demands. Nevertheless, continuous monitoring and management measures are imperative to ensure the sustained and efficient operation of these locations. Implementing dynamic adjustments based on usage patterns and real-time demand fluctuations can further optimize the overall operational efficiency of lightly congested spots.
\end{itemize}

\begin{figure}[ht]
    \begin{center}
        \includegraphics[width=0.8\linewidth]{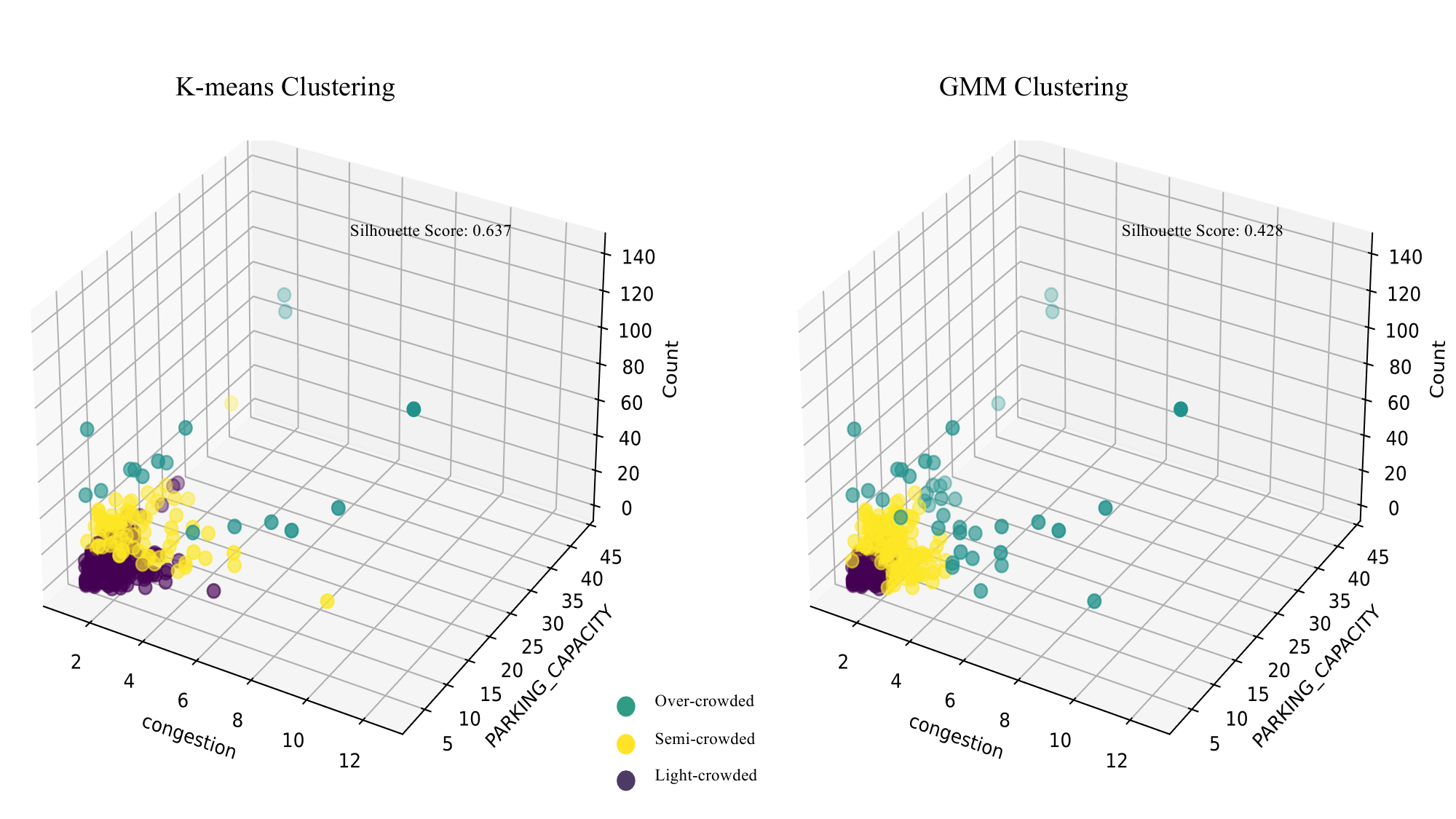}
    \end{center}
    \caption{BS congested spots classification result}
    \label{fig:cluster}
\end{figure}

\begin{figure}[ht]
    \begin{center}
        \includegraphics[width=0.8\linewidth]{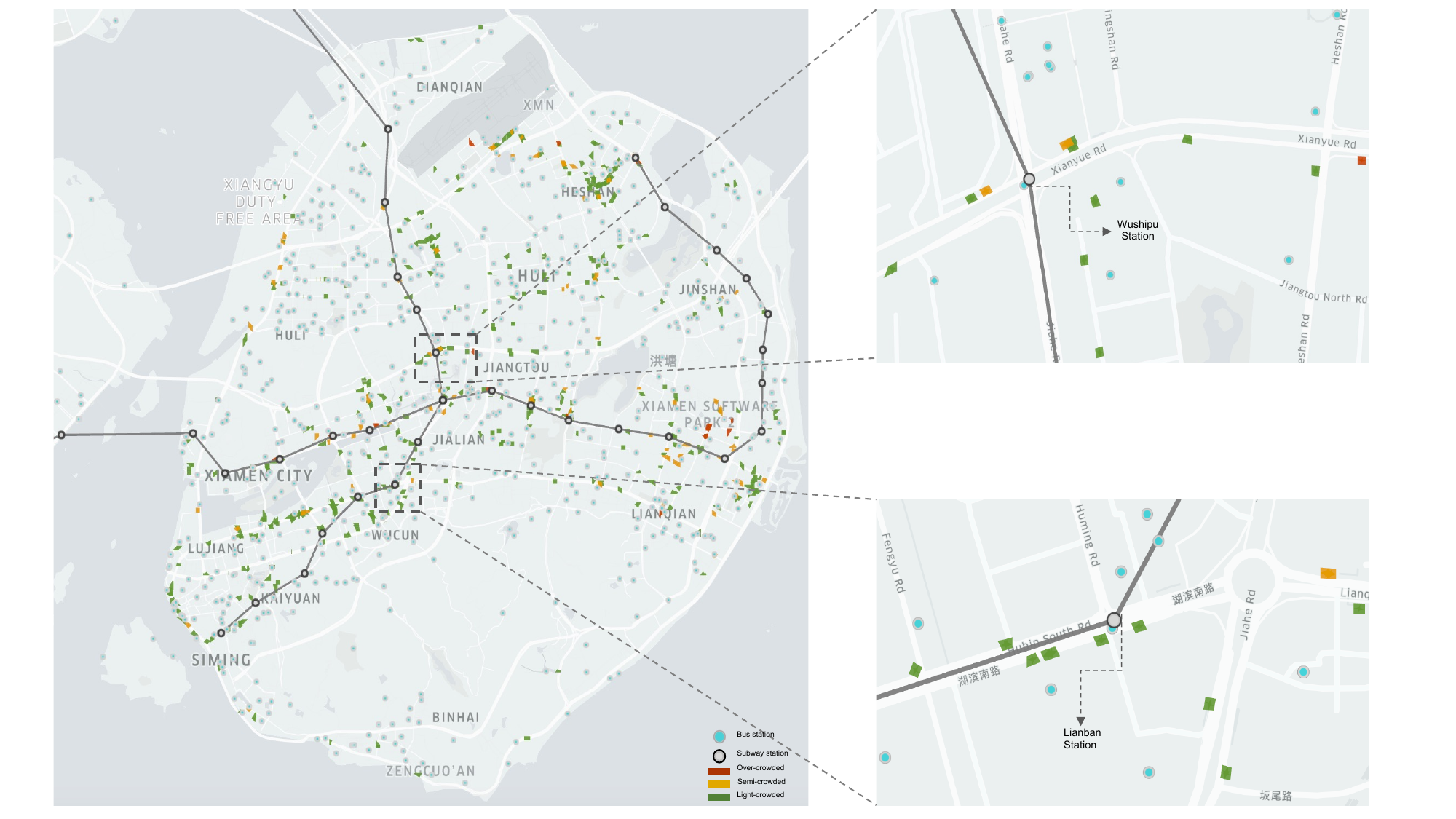}
    \end{center}
    \caption{Result of classifying congested BS parking spots.}
    \label{fig:spots_map}
\end{figure}

Fig.~\ref{fig:spots_map} illustrates the spatial distribution of the classified congested parking spots for bike-sharing (BS). The increased concentration of commuters around these transportation hubs contributes to the observed congestion in the shared bicycle parking spots. However, this dense distribution, while meeting the demand at these key locations, may inadvertently lead to resource dispersion and reduced utilization rates for each station. Users tend to prefer stations closer to their origin or destination points, potentially neglecting nearby stations, even in popular districts. This phenomenon underscores the need for nuanced strategies to optimize station placement, enhance user accessibility, and improve overall service efficiency during peak hours.

\section{Conclusions and Future Directions}\label{sec:Conclusions}

In conclusion, this study provides a comprehensive understanding of urban daily activity patterns through the innovative identification and classification of congested bike-sharing (BS) parking spots within the context of Xiamen's bike-sharing system. Our framework integrates the Congestion Density($C$) metric and the K-means algorithm, showcasing the potential of data-driven decision-making in urban planning and bike-sharing management. By focusing on individual BS parking spots rather than broader clusters, we offer a nuanced perspective on the congested parking situation in Xiamen, identifying and classifying 563 congested spots into Over-crowded, Semi-crowded, and Light-crowded clusters.

The key contributions and findings of this study are summarized as follows:

\begin{itemize}
\item (i) Detailed insights into urban mobility patterns, including commuter behaviors, trip distances, and durations, underscore the significance of bike-sharing for short urban trips.
\item (ii) Identification of the alignment between established subway lines, bus stations, and bike-sharing routes, emphasizing the popularity of the "subway + bike-sharing" and "bus + bike-sharing" paradigm, particularly during morning peak hours.
\item (iii) Quantitative identification of congested parking spots at different time intervals using the Congestion ($C$) metric.
\item (iv) Classification of congested parking spots into Over-crowded, Semi-crowded, and Light-crowded categories using the K-means method, providing actionable insights for resource allocation and bike-sharing system optimization.
\end{itemize}

While the study yields promising outcomes, several avenues for further research emerge:

\begin{itemize}
\item (i) \textbf{Cross-City Comparisons:} Explore the applicability of the identified congestion patterns and classification framework in other cities. Conduct cross-city comparisons to understand variations in bike-sharing congestion dynamics, considering factors such as city layout, transportation infrastructure, and cultural influences.
\item (ii) \textbf{In-Depth User Behavior Analysis:} Extend the analysis to delve deeper into user behaviors. Investigate factors influencing bike-sharing choices, preferred drop-off points, and user preferences during congested periods. This deeper understanding can inform user-centric strategies and enhance the overall user experience.
\item (iii) \textbf{Data Integration:} Enrich the dataset by incorporating additional relevant data sources, such as weather conditions, special events, and socio-economic factors. This expanded data integration can provide a more holistic view of the factors influencing bike-sharing congestion and contribute to the development of robust predictive models.
\item (iv) \textbf{Dynamic Capacity Management:} Investigate the feasibility of dynamic capacity adjustments for bike-sharing parking spots based on real-time demand. Implementing adaptive capacity management strategies can enhance system efficiency and responsiveness to fluctuating congestion levels.
\end{itemize}

Addressing these future directions will contribute to a more comprehensive understanding of urban mobility dynamics and refine strategies to enhance bike-sharing services and urban planning on a broader scale.

\section*{Acknowledgments}
\addcontentsline{toc}{section}{Acknowledgments}
This work was supported by the National Natural Science Foundation of China, Youth Science Fund Project (Grant No.11901097). The authors express their heartfelt gratitude to the editor and reviewers for their invaluable and insightful comments.

\clearpage

\printcredits

\bibliographystyle{apalike}

\bibliography{manuscript/refs}

\bio{}
\endbio

\end{document}